\documentclass[aps,prd,twocolumn,prd,showpacs,showkeys,preprintnumbers,bibnotes,floatfix,longbibliography,notitlepage,nofootinbib,superscriptaddress,superscriptgaddress,
]{revtex4-2}

\pdfoutput=1
\usepackage{amsmath}
\usepackage{amsfonts}
\usepackage{amssymb}
\usepackage{mathrsfs}
\usepackage{graphicx}
\usepackage{color}
\usepackage{bbding}
\usepackage{tabularx}
\usepackage[usenames,dvipsnames,table]{xcolor}
\usepackage[normalem]{ulem}
\usepackage{makecell}
\usepackage{slashed}

\usepackage[hidelinks]{hyperref}
\usepackage{multirow}
\usepackage{makecell}
\usepackage{upgreek}
\usepackage[capitalise]{cleveref}
\usepackage{subfigure}
\usepackage[normalem]{ulem}

\begin{document}

\title{Supernova cooling from neutrinophilic dark matter}
\author{Yugen Lin}
\email{linyugen@itp.ac.cn}
\affiliation{Institute of Theoretical Physics, Chinese Academy of Sciences, Beijing, 100190, China}

\begin{abstract}
Core-collapse supernova serve as a powerful laboratory for testing physics beyond the Standard Model (BSM), particularly regarding new, light states interacting feebly with SM particles. In this work, we investigate
for the first time the production of dark matter (DM) via the neutrino-neutrino scattering processes
inside a core-collapse supernova, which contributes to the excessive cooling.  By incorporating state-of-the-art supernova simulation data , we derive stringent
and robust limits on sub-GeV dark matter with effective couplings to neutrinos. We find that the existing and projected constraints from indirect detection are quite weak. Our supernova cooling bounds on DM-neutrino reference cross section can improve indirect detection limits more than ten orders of magnitude for DM masses below $\mathcal{O}(100)$~MeV, and it can also provide strong complementarity with other cosmological constraints. Our results highlight the exceptional sensitivity of core-collapse supernova to feebly interacting particles and motivate future supernova neutrino observations as a powerful probe of light dark sectors.

\end{abstract}

\maketitle

\section{Introduction}
The nature of dark matter (DM) remains one of the
most profound puzzles in modern physics. For the known Standard Model sector, we also lack a complete understanding for the properties of neutrinos. Dark matter and neutrino are the two most compelling pieces for the existence of new physics beyond the Standard Model (BSM), and both of them remain largely unknown. Therefore, it is of great interest to investigate whether the two sectors are intimately
related.

From an experimental perspective, probing a possible connection between dark matter and neutrinos is challenging, as both species are essentially invisible to detectors. However, such interactions are not entirely hidden, and they inevitably leave imprints on a wide range of cosmological and astrophysical observables, such as Cosmic Microwave Background (CMB)~\cite{Mangano:2006mp,Wilkinson:2014ksa,Brax:2023tvn}, the formation of small-scale structure~\cite{Boehm:2000gq,Boehm:2004th}, and galactic density profiles~\cite{Akita:2023yga,Heston:2024ljf}, etc. A comprehensive overview of DM-neutrino interaction can be found in recent paper.~\cite{Dev:2025tdv}.

On the other hand, core-collapse supernova is also an ideal astrophysical environment for investigating DM-neutrino interactions. The hot and dense core of supernova facilitates the efficient production of exotic particles up to $\mathcal{O}(100)$~MeV mass, even if the particle is only weakly coupled. These particles may escape from the supernova, introducing an additional cooling channel for the protoneutron star and altering the emissivity of neutrinos~\cite{Raffelt:2006cw}. Since the detection of the neutrino burst from Supernova SN1987A~\cite{Kamiokande-II:1987idp}, supernova cooling has been employed to probe various BSM models (see e.g.~\cite{Ellis:1987pk,Raffelt:1987yt,Raffelt:1996wa,Chang:2018rso,Fiorillo:2022cdq,Li:2025beu,Cappiello:2025tws,Caputo:2025aac,Fiorillo:2023ytr}.)

In our recent work~\cite{Lin:2025mez}, we investigate the production of dark matter in Supernova via neutrino-electron/nucleon scattering (i.e. $\nu + e^-/N \rightarrow \chi + e^-/N$) for the first time. A natural and yet unexplored channel is the neutrino–neutrino scattering process $\nu+\nu \rightarrow \chi+\chi$ where two SM neutrinos scatter into a pair of DM particles. In this paper, we fill this gap by performing the first systematic analysis of supernova constraints on the neutrino–neutrino scattering into DM. Besides, we also perform the analysis for DM production via neutrino annihilation process $\nu+\bar{\nu} \rightarrow \chi+\bar{\chi}$. We consider both scalar and vector interactions, calculate the corresponding freeze-in overproduction limits, and compare our cooling bounds with other existing indirect detection constraints.

The paper is organized as follows. In Sec.~II we describe the effective Lagrangian for the neutrino-DM interaction and present the relevant UV realization. Sec.~III and IV detail the supernova modeling, the DM production, propagation, and the cooling criterion. Our main results, including the excluded parameter regions and comparisons with other bounds, are shown in Sec.~V, and we conclude in Sec.~VI.

\section{MODEL SETUP}
DM-neutrino interaction can be studied both from the UV theory and effective field theory. To make a systematic study, we take the effective field theory
approach for a model independent analysis. Here we assume DM dominantly couples only to neutrinos and no other SM species at leading order. Therefore, the relevant degrees of freedom only include neutrino and fermionic DM particle. This is theoretically possible, for instance, if the DM couplings to charged particles in the SM are suppressed either by mixing, higher-order effects, or higher-dimensional operators~\cite{Blennow:2019fhy,Farzan:2016wym,Ma:2006km}.

\begin{figure}[!htbp]
\centering
\includegraphics[width=\columnwidth]{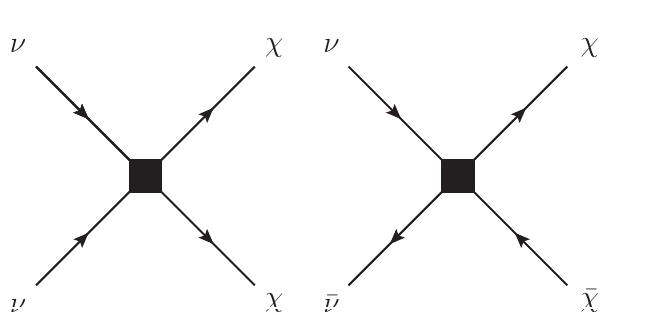}
\caption{Feynman diagrams of DM production for the interactions in Eqs.~\eqref{eq:vectoroperator}~-~\eqref{eq:s-vectoroperator}, referred to as scattering (left) and annihilation (right) processes, respectively.}
\label{fig:Feynman diagrams}
\end{figure}

The leading order interactions for DM-neutrino interaction are dimension-six operators. We consider two scenarios: DM interacts with neutrino via scattering and annihilation processes. For the former, the most commonly studied operators are vector- and scalar-type operators given by
\begin{align}
&O_V^{sca} =\frac{1}{\Lambda^2}\left(\bar{\chi} \gamma^\mu \nu_L \right)\left(\bar{\nu}_L \gamma_\mu \chi\right)\,,
\label{eq:vectoroperator}\\
&O_S^{sca} =\frac{1}{\Lambda^2}\left(\bar{\chi} \nu_L\right)(\bar{\nu}_L \chi)\,,
\label{eq:scalaroperator}
\end{align}
where $1/\Lambda^2$ is the Wilson coefficient determined by the heavy mediator mass and its coupling in the UV complete model. The neutrino field is taken to be the SM left-handed component. The latter can be written as
\begin{align}
&O_V^{ann} =\frac{1}{\Lambda^2}\left(\bar{\chi} \gamma^\mu \chi \right)\left(\bar{\nu}_L \gamma_\mu \nu_L\right)\,.
\label{eq:s-vectoroperator}
\end{align}
Due to the absence of right-handed neutrino in the Standard Model, here we don't consider the scalar-type DM-neutrino interaction in the annihilation process. The Feynman diagrams of DM production via neutrino-neutrino scattering and neutrino-antineutrino annihilation processes are shown in Fig.~\ref{fig:Feynman diagrams}.

The above effective operators are naturally realized in many UV models. For example, for the coupling in the form of $\phi/Z^\prime\nu\chi$,  a representative UV-completion of such operator appears in the radiative generation of neutrino masses at loop level~\cite{Arhrib:2015dez,Babu:2019mfe,Herms:2023cyy}, the corresponding Lagrangian can be written as
\begin{equation}
    \mathcal{L}\supset  g_{\chi\nu}\bar{\chi}\gamma_\mu P_L\nu Z'^\mu\,+ \rm{h.c.},
\end{equation}

\begin{equation}
    \mathcal{L}\supset  g_{\chi\nu}\bar{\chi} P_L\nu \phi\,+ \rm{h.c.},
\end{equation}
where $Z'^\mu$ and $\phi$ are vector/scalar fields respectively. In this work, we consider the DM-neutrino interactions are generated via heavy mediators. By integrating out these heavy fields, we can obtain the effective operator in Eqs.~\eqref{eq:vectoroperator}~-~\eqref{eq:scalaroperator}. For the interaction in Eq.~\eqref{eq:s-vectoroperator}, it can be achieved with the following Lagrangian~\cite{Abdallah:2021npg,Nomura:2017wxf}
\begin{equation}
    \mathcal{L}\supset g_\nu\bar{\nu}\gamma_\mu P_L\nu Z'^\mu + g_\chi\bar{\chi}\gamma_\mu \chi Z'^\mu\,.
\end{equation}
Similarly, after integrating out the heavy field  $Z'^\mu$, we can obtain the operator in Eq.~\eqref{eq:s-vectoroperator}. Given that the mediator masses are much larger than the energy of supernova neutrinos, the different UV models will not affect the supernova cooling constraints.

\begin{figure*}[!htbp]
\centering
\includegraphics[width=\columnwidth]{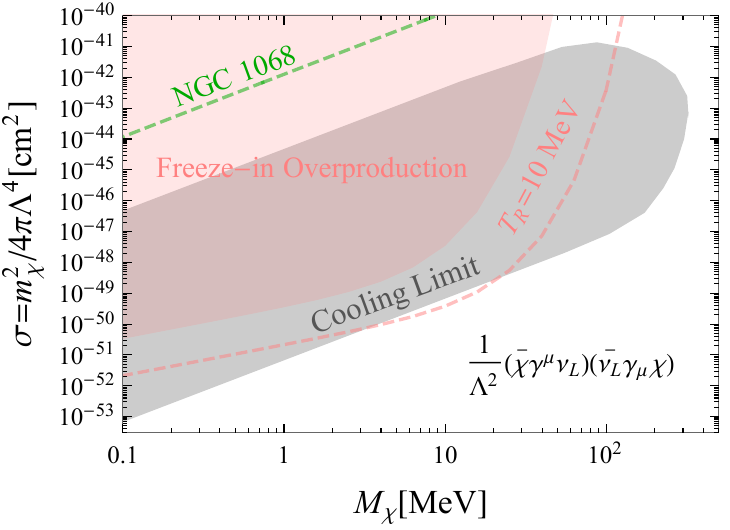}
\includegraphics[width=\columnwidth]{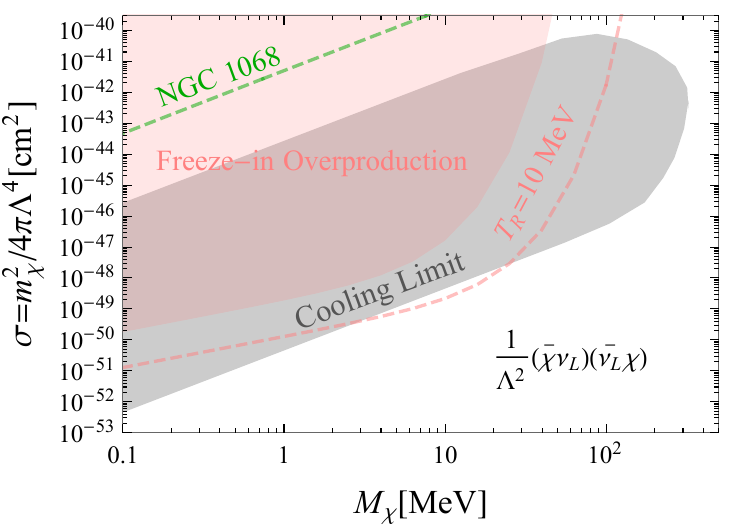}
\caption{Supernova cooling limits (gray shaded regions) for neutrino scattering processes via vector-type (left) and scalar-type interactions (right). The constraints from DM overproduction via freeze-in are shown in pink regions, and the pink dashed lines represent the freeze-in processes with a higher reheating temperature $T_R=10$ MeV. The constraints from high-energy neutrino attenuation with DM spikes are shown in dashed green lines.}
\label{fig:coolinglimit1}
\end{figure*}

\section{SUPERNOVA COOLING}
In a core-collapse supernova, neutrinos are abundantly produced via various processes, including neutronization, beta decay, and electron-positron annihilation. These neutrinos can scatter and annihilate on their way of propagation outside the supernova, producing DM efficiently via the effective interaction in Eqs.~\eqref{eq:vectoroperator}~-~\eqref{eq:s-vectoroperator}.
The corresponding total cross sections for neutrino scattering and annihilation are given by
\begin{equation}
\sigma^{V(sca)}_{\nu_a \nu_a \rightarrow \chi \chi}=
\frac{1}{8\pi \sqrt{s}\Lambda^4}\sqrt{s-4m_\chi^2}\left(s-2m_\chi^2\right), \label{eq:sigmavector}
\end{equation}

\begin{equation}
\sigma^{S(sca)}_{\nu_a \nu_a \rightarrow \chi \chi}=
\frac{1}{48\pi \sqrt{s}\Lambda^4}\sqrt{s-4m_\chi^2}\left(s-m_\chi^2\right), \label{eq:sigmascalar}
\end{equation}

\begin{equation}
\sigma^{V(ann)}_{\nu_a\bar{\nu}_a \rightarrow \chi\bar{\chi}}=
\frac{1}{6\pi \sqrt{s}\Lambda^4}\sqrt{s-4m_\chi^2}\left(s+2m_\chi^2\right), \label{eq:sigmavector-ann}
\end{equation}
where the subscript $a$ represents neutrino species, and s is the squared center-of-mass (COM) energy.

When dark matter particles interact weakly, they can freely stream out of  the supernova core. The energy loss rate per unit volume by scattering $\nu\nu \rightarrow \chi\chi$ or annihilation  process $\nu\bar{\nu} \rightarrow \chi\bar{\chi}$ is

\begin{equation}
\begin{aligned}
Q= & \int\left[\prod_{i=1}^4 \frac{d^3 \vec{p}_i}{(2 \pi)^3 2 E_i}\right](2 \pi)^4 \delta^4\left(p_1+p_2-p_3-p_4\right) \\
& \times f_1 f_2\left(1-f_3\right)\left(1-f_4\right) \sum_{\text {spins }}|\mathcal{M}|^2 \left(E_1+E_2\right).
\label{eq:energyloss}
\end{aligned}
\end{equation}
During the dark matter production process, $f_\chi \ll f_\nu$, we can safely omit the DM factor $1-f_{3(4)}$. Performing the integral of final state phase space, we obtain

\begin{equation}
Q=\int \frac{d^3 p_1}{(2 \pi)^3 2 E_1} \frac{d^3 p_2}{(2 \pi)^3 2 E_2} f_1 f_2\left(E_1+E_2\right) 4 E_1 E_2 \sigma v_{\mathrm{rel}}
\end{equation}

Under spherical symmetry, the above energy loss rate can be simplified as
\begin{equation}
Q=\frac{1}{4 \pi^2} \int \mathrm{d} E_1 \mathrm{d} E_2 \mathrm{d} \cos \theta F_1 F_2 \left(E_1+E_2\right)\sigma v_{\mathrm{rel}}
\label{eq:energylossrate}
\end{equation}
where $E_1$, $E_2$ are the energy of two initial state neutrino respectively, $v_{\mathrm{rel}}=\sqrt{2\left(1-\cos\theta\right)}$ is the relative velocity between the two neutrinos. $F_{1,2}$ represent the neutrino energy spectrum,
\begin{equation}
F_{1,2}\equiv \frac{d n_\nu}{d E_\nu}=\frac{n_\nu}{\bar{E}_\nu}f_\nu\left(E_\nu\right),
\label{eq:energy spectrum}
\end{equation}
$n_\nu$ is the neutrino number density, $\bar{E}_\nu$ is the mean neutrino energy, and $f_\nu$ is the neutrino distribution function,
\begin{equation}
f_\nu=\frac{(1+\alpha)^{(1+\alpha)}}{\Gamma(1+\alpha)}\left(\frac{E_\nu}{\bar{E}_\nu}\right)^\alpha \operatorname{Exp}\left[-\left(1+\alpha\right) \frac{E_\nu}{\bar{E}_\nu}\right]\,,
\label{eq:distribution function}
\end{equation}
$\alpha$ is a fit parameter, which sometimes referred to as $\alpha$-fit~\cite{Keil:2002in}, and its typical values lie in the range $\alpha \sim 2-5$. $n_\nu$, $\alpha$ and $\bar{E}_\nu$ are a function of time and space explicitly.

For a given physical process, the energy loss rate Eq.~\ref{eq:energylossrate} can be further simplified. We can write the angular dependence explicitly and analytically perform the angular integral. The angular dependence in Eq.~\ref{eq:energylossrate} is through relative velocity $v_{\rm rel}$ and the squared COM energy $s=2E_1E_2(1-\cos\theta)$. After integrating out the angular part, we can obtain the emissivity
formula
\begin{equation}
Q=\frac{1}{4 \pi^2} \int \mathrm{d} E_1 \mathrm{d} E_2 F_1 F_2 \left(E_1+E_2\right) L(E_1,E_2,m_\chi),
\label{eq:energylossrate1}
\end{equation}
where the function $L$ for different processes are respectively
\begin{equation}
L^{V}_{sca}=\frac{2}{15\pi \Lambda^4}\left(6E_1E_2-m_\chi^2\right)\left(1-\frac{m_\chi^2}{E_1E_2}\right)^{3/2},
\end{equation}

\begin{equation}
L^S_{sca}=\frac{1}{30\pi \Lambda^4}\left(4E_1E_2+m_\chi^2\right)\left(1-\frac{m_\chi^2}{E_1E_2}\right)^{3/2},
\end{equation}

\begin{equation}
L^V_{ann}=\frac{1}{5\pi \Lambda^4}\left(2E_1E_2+3m_\chi^2\right)\left(1-\frac{m_\chi^2}{E_1E_2}\right)^{3/2}.
\end{equation}
We sum over all flavor species of neutrinos and antineutrinos to obtain the total energy loss rate.

We extract the space-time dependent variables neutrino number density $n_\nu$, spectrum parameter $\alpha$, and mean neutrino energy $\bar{E}_\nu$ from an 8.8$M_\odot$ progenitor star simulated by the Garching group~\cite{Hudepohl:2009tyy}. In our previous analysis~\cite{Lin:2025mez}, we show the results are robust against the choice of supernova profiles and are conservative compared with models with higher progenitor masses. The dark matter luminosity is $L_\chi=\int Q dV$, and the volume integral extends from the supernova center (close to $R^\prime = 0$) to its outer layers, and we stop at 40 km in our calculation. This cutoff is reasonable since the production of DM predominantly occurs within the neutrino sphere ($\sim$ 30 km), where the number densities of neutrino are orders of magnitude larger than outside.

\begin{figure}[!htbp]
\includegraphics[width=\columnwidth]{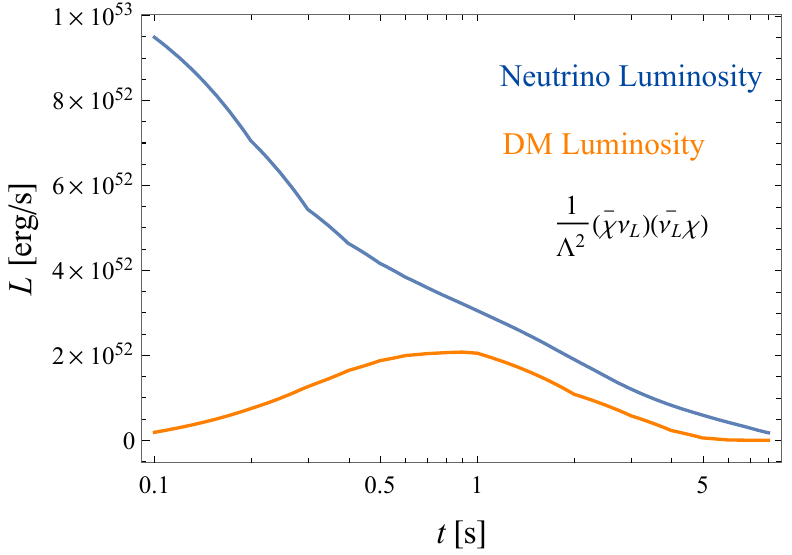}
\caption{Time evolution of the DM luminosity (orange curve) and
neutrino luminosity (blue curve). Here we choose a scalar-type interaction in Eq.~\ref{eq:scalaroperator} with $M_\chi=1$ MeV and $\Lambda=10~\rm{TeV}$ for illustration.}
\label{fig:DMluminosity}
\end{figure}

The upper limits on DM–neutrino interactions are derived from the Raffelt criterion~\cite{Raffelt:1990yz}, which requires that the luminosity carried by any novel particles escaping the protoneutron star environment must not exceed the neutrino luminosity. Otherwise, an additional efficient cooling channel would have shortened the supernova neutrino burst duration relative to the observed signal from SN1987A. Here we use $L_\chi \lesssim L_\nu=3\times10^{52}$ erg/s at one second post-bounce to obtain the cooling limits, which is consistent with the neutrino emission signal detected during SN 1987A. In Fig.~\ref{fig:DMluminosity},  we show the time evolution of the DM luminosity (orange curve) for $M_\chi=1$ MeV and $\Lambda=10~\rm{TeV}$, and the neutrino luminosity (blue curve) extracted from the simulation data~\cite{Hudepohl:2009tyy}.  Because the DM luminosity peaks around 1~s after bounce, a luminosity comparison at that time (i.e. Raffelt criterion) captures the critical feature of DM cooling over the period of the supernova evolution.

\section{DARK MATTER PROPAGATION}
We now estimate the onset of trapping and identify where free-streaming bounds cease to apply for the cooling constraints. For produced DM particles, if $1/\Lambda^2$ is small enough, virtually all of the DM will escape without scattering.  However, If $1/\Lambda^2$ is too large, the produced DM particle may further scatter with the neutrino and even be trapped in the supernova. Conservatively, in this work, we restrict ourselves to DM that escapes without scattering.

We can calculate the survival probability of DM
\begin{equation}
P\left( E_\chi,r\right)=\mathrm{Exp} \left(-\int_{r}^{\infty} \frac{\mathrm{d} r'}{\lambda\left(E_\chi,r'\right)}\right),
\label{eq:escape probability}
\end{equation}
where $\lambda\left(r\right)$ is the mean free path of DM. For DM-neutrino scattering,
\begin{equation}
\lambda^{-1}(E_\chi,r)
=\frac{n_{\nu}(r)}{2v_\chi}
\int_{-1}^{1}d\cos\theta\,
\sigma_{\chi \nu \rightarrow \chi\nu }^{\rm PB} v_{\rm rel}.
\label{eq:mfp}
\end{equation}
where $v_{\rm rel}=\sqrt{1-v_\chi\cos\theta+v_\chi^2}$ is the relative velocity between DM and neutrino, $n_\nu$ is the neutrino number density. The final state neutrino will experience a blocking effect as some of the phase space is already occupied by existing neutrinos with a substantial number density. The cross section in mean free path need to be corrected by including the Pauli blocking effect,
\begin{equation}
\sigma_{\chi \nu \rightarrow \chi\nu }^{\rm PB}
=
\int d\Omega_*\,
\frac{d\sigma_{\chi \nu \rightarrow \chi\nu }}{d\Omega_*}
\left[
1-f_{\rm FD}
\left(
E_\nu',\mu,T
\right)
\right],
\label{eq:sigma_PB}
\end{equation}
where $d\sigma/d\Omega_*$ is defined in the center-of-momentum frame, $f_{\rm FD}$ is the Fermi-Dirac distribution of neutrinos and $E_\nu^\prime$ is the final state neutrino energy in the supernova frame. Here we assume that DM particles produced at radius $r$ travel on radial trajectories out of the supernova. The average energy of DM produced at $r$ can be estimated as $\bar{E}_{\rm \chi}=Q_\chi/\dot{n}_\chi$, where $\dot{n}_\chi$ can be obtained from Eq.~\ref{eq:energylossrate} without the $(E_1+E_2)$ factor. In Appendix.~\ref{sec:simulationdata}, we show these parameters from supernova simulation data~\cite{Hudepohl:2009tyy}, including the neutrino number density $n_\nu$, mean neutrino energy $\bar{E}_\nu$,  electron neutrino chemical potential $\mu$, and supernova temperature $T$ at one second post-bounce. We then multiply the survival probability by the Eq.~\ref{eq:energylossrate} and integrate over space to calculate the trapping-corrected energy loss rate by
DM. The trapping limits can be obtained when the interaction is strong enough such that the Raffelt criterion is met again.

Altogether, we find supernova cooling sets stringent constraints on DM-neutrino interaction, which span by about seven orders of magnitude in the cross section, regardless of the processes and the type of interactions. It rules out the cross sections as low as $10^{-46}-10^{-53}$ cm$^2$ for neutrino scattering and $10^{-45}-10^{-52}$ cm$^2$ for neutrino annihilation in the $0.1–100$ MeV DM mass range. Due to the low number density of antineutrino in the supernova, it makes the cooling constraints of neutrino-antineutrino annihilation is weaker than constraints from neutrino-neutrino scattering. The above limits can be extended to arbitrarily
smaller masses of DM particles.

It should be emphasized that the trapping limits we derive here are conservative, which roughly corresponds to the mean-free paths comparable to the supernova core radius. In reality, DM is only fully trapped at lower mean-free-path with larger cross section, which would further broaden our exclusion region.

\section{DARK MATTER OVERPRODUCTION AND OTHER CONSTRAINTS}
For cosmological dark matter, we need to calculate the corresponding relic density. Requiring the DM relic
density to be less than the measured value, $\Omega_\chi h^2 \lesssim 0.12$, sets a upper bound on DM-neutrino cross section. In other words, the DM relic density cannot be overproduced.

Light DM typically has a very small coupling with SM
particles, which makes it difficult for the light DM to reach thermal equilibrium with the environmental plasma. Therefore, their production is usually realized by the freeze-in mechanism~\cite{Hall:2009bx,DEramo:2017ecx,Gopalakrishna:2006kr,Page:2007sh}.  For the effective operators, the freeze-in contribution to the DM abundance is UV dominated, so that the DM abundance will grow with the reheating temperature $T_R$~\cite{Manzari:2025jbc,Lehmann:2020lcv}. Consequently, to place conservative constraints on DM-neutrino cross section from DM overproduction, we use the lowest allowed reheating temperature $T_R=4$ MeV~\cite{Kawasaki:2000en,Hannestad:2004px}.

\begin{figure}[!htbp]
\centering
\includegraphics[width=\columnwidth]{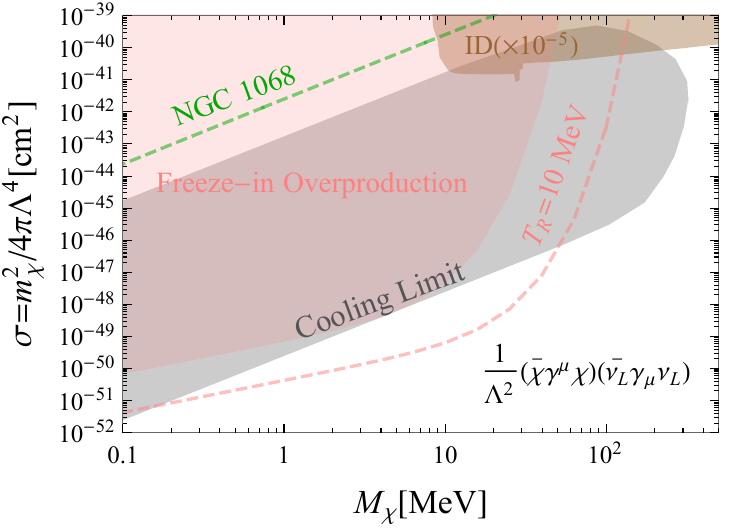}
\caption{Same as Fig.~\ref{fig:coolinglimit1}, but for neutrino annihilation process via vector-type interaction. The brown shaded region shows the combined indirect detection constraints, with the
limits multiplied by $10^{-5}$ for visualization.}
\label{fig:coolinglimit2}
\end{figure}

For the effective operator Eq.~\eqref{eq:vectoroperator} and Eq.~\eqref{eq:scalaroperator}, DM can be produced via processes $\nu\nu \rightarrow \chi\chi$, $\nu\bar{\nu} \rightarrow \chi\bar{\chi}$, $\bar{\nu}\bar{\nu} \rightarrow \bar{\chi}\bar{\chi}$ in the early Universe, and for the operator
Eq.~\eqref{eq:s-vectoroperator}, DM is only produced via process $\nu\bar{\nu} \rightarrow \chi{\bar\chi}$. The Boltzmann equation govern the evolution of DM number density $n_\chi$ in the early Universe, which can be expressed as

\begin{equation}
\begin{aligned}
&\frac{d n_\chi}{dt}+3Hn_\chi= \int\left[\prod_{i=1}^4 \frac{d^3 \vec{p}_i}{(2 \pi)^3 2 E_i}\right](2 \pi)^4 \delta^4\left(p_1+p_2-p_3-p_4\right) \\
& \times |\mathcal{M}|^2 \left[f_1 f_2\left(1-f_3\right)\left(1-f_4\right)- f_3 f_4\left(1-f_1\right)\left(1-f_2\right)\right].
\label{eq:Boltzmann}
\end{aligned}
\end{equation}
where the subscript 1, 2 and 3, 4 represent initial state neutrinos and final state dark matter respectively, $f_i$ is its phase space distribution function, and $H$ is the Hubble parameter.

\begin{figure}[!htbp]
\includegraphics[width=\columnwidth]{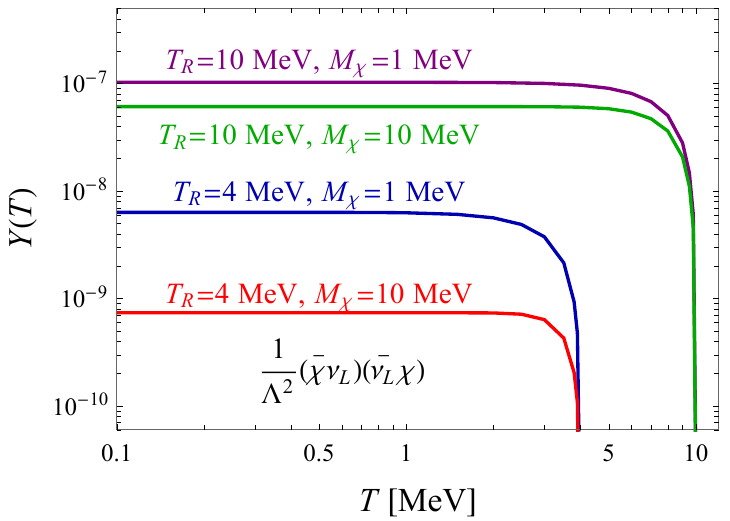}
\caption{The evolution of the DM yield $Y(T)$ as function of temperature $T$. Here we choose a scalar-type interaction in Eq.~\ref{eq:scalaroperator} with $\Lambda=10~\rm{TeV}$ for illustration. The different colored lines represents different reheating temperature and the mass of DM particle.}
\label{fig:YT}
\end{figure}

With the freeze-in mechanism, the DM density increases from 0. During the DM production process $f_\chi \ll f_\nu$. Thus, the second term of Eq.~\ref{eq:Boltzmann} can be omitted. Introducing the DM yield $Y=n_\chi/s(T)$, $s(T)$ is entropy density. At the epoch of
DM production, the Universe is dominated by radiation. The Hubble parameter $H$ and entropy density $s(T)$ are $H=1.66\sqrt{g_\star}\frac{T^2}{M_p}$, $s(T)=\frac{2\pi^2}{45}g_{\star s}T^3$, where  $g_\star$, $g_{\star s}$ respectively represent relativistic degrees of freedom of energy and entropy densities, and $M_p=1.22 \times 10^{19}$ GeV is the Planck mass. Then in terms of DM yield Y, the Boltzmann equation becomes

\begin{equation}
\frac{d Y}{d T}=-\frac{45 M_{\mathrm{P}}}{2 \pi^2\left(1.66 \sqrt{g_*}\right) \tilde{g}_{* s} T^6}\left\langle \sigma_{1,2 \rightarrow3,4} v \right\rangle n_{1}^{\mathrm{eq}} n_{2}^{\mathrm{eq}}
\label{eq:BEY}
\end{equation}
where $\tilde{g}_{* s}=g_{\star s}\left(1+\frac{T}{3 g_{\star s}}\frac{dg_{\star s}}{dT}\right)$. In our interested region, we take $g_\star= g_{\star s}\simeq 10.75$. Following Ref.~\cite{Gondolo:1990dk}, the thermally averaged cross section can be simplified as

\begin{equation}
\left\langle \sigma v\right\rangle n_{1}^{\mathrm{eq}} n_{2}^{\mathrm{eq}}=\frac{T}{8 \pi^4} \int d s\left(s-4 m_i^2\right) \sqrt{s} K_1\left(\frac{\sqrt{s}}{T}\right) \sigma(s),
\end{equation}
where $K_1$ is the first modified Bessel function of second kind, the initial state neutrino mass $m_i=0$, and $s \geq \rm max (4m_i^2, 4m_f^2)$. The solution of the Boltzmann equation can be obtained by integrating the temperature $T$ from $T_R$

\begin{equation}
\begin{aligned}
Y(T) &= \frac{45 M_{\mathrm{P}}}{16 \pi^6} \int_T^{T_{R }} \frac{d \widetilde{T}}{\left(1.66 \sqrt{g_*}\right) {g}_{* s} \widetilde{T}^5}\\
&\int_{4 m_\chi^2}^{\infty} d s\left(s-4 m_i^2\right) \sqrt{s} K_1\left(\frac{\sqrt{s}}{\widetilde{T}}\right) \sigma(s),
\label{eq:BE}
\end{aligned}
\end{equation}
Here we sum over all physical processes of DM production to obtain the DM yield today $Y(0)$. In Fig.~\ref{fig:YT}, we show the evolution of the DM yield $Y(T)$ as function of temperature $T$. Taking the scalar-type interaction in Eq.~\ref{eq:scalaroperator} with $\Lambda=10~\rm{TeV}$ as an example, we choose two different reheating temperature $T_R=4$ MeV and 10 MeV, and two different DM mass $M_\chi=1$ MeV and 10 MeV, the corresponding evolution functions are shown as blue, red, purple, and green solid lines respectively.

The DM relic density is estimated as
\begin{equation}
\Omega_\chi h^2=\frac{2m_\chi Y_0s_0h^2}{\rho_c}
\end{equation}
where $s_0=2970$ cm$^{-3}$ is the present entropy density, $\rho_c=1.054\times10^{-5}h^2$GeV cm$^{-3}$ is the critical density, and $h = 0.67$ is the reduced Hubble constant. The factor 2 in the above formula is from that both DM and its anti-particle can be produced in the early Universe. Using $\Omega_\chi h^2 \lesssim 0.12$, we can obtain the upper limits on DM-neutrino cross section.

We show our cooling limits for neutrino scattering and neutrino annihilation processes in Fig.~\ref{fig:coolinglimit1} and Fig.~\ref{fig:coolinglimit2} respectively, which are shown in gray shaded regions. We define the reference DM-neutrino cross section $\sigma \equiv m_\chi^2/(4\pi\Lambda^4)$ to present the corresponding constraints. For cosmological DM, the corresponding overproduction constraints are shown in pink regions. We also show the overproduction constraints with a higher reheating temperature $T_R=10$ MeV, which is shown in pink dashed line. The green dashed line represents the excluded parameter space derived from the absence of high-energy neutrino attenuation from the nearby active galaxy NGC 1068~\cite{Cline:2023tkp}, under the assumption of an existing DM spike. Indirect detection can also set constraints on DM-neutrino cross section from DM annihilation processes using neutrino detectors. We compare limits from analyses of Borexino~\cite{Borexino:2010zht,Borexino:2010dli}, KamLAND~\cite{KamLAND:2011bnd,KamLAND:2013rgu}, and Super-K~\cite{Super-Kamiokande:2008ecj,Super-Kamiokande:2010tar} data, and the projection of Hyper-K~\cite{Bell:2020rkw} and JUNO experiments~\cite{JUNO:2015zny}, where we find the maximum sensitivity occurs at $m_\chi=30$ MeV and the corresponding reference cross section is around $10^{-36}$ cm$^2$, which is far above our cooling limits. For illustration, we take the vector annihilation operator
$\mathcal{O}^{\rm ann}_V$ as a example and recast
the indirect-detection constraints compiled in
Ref.~\cite{Arguelles:2019ouk} as the brown shaded region in Fig.~\ref{fig:coolinglimit2}.
To bring these constraints into the displayed range, we rescale the corresponding upper limits by a factor of $10^{-5}$. Indirect searches rely on neutrinos from DM annihilation in subhalos. In a supernova core, the hot and dense neutrino bath provides an efficient source of DM through $\nu\bar{\nu}\to\chi\bar{\chi}$, which cause the much stronger sensitivity comparing with the indirect-detection constraints. While cosmological bounds, such as those from the Cosmic Microwave Background~\cite{Mangano:2006mp,Wilkinson:2014ksa} and small-scale structure formation~\cite{Boehm:2000gq,Boehm:2004th,Akita:2023yga,Heston:2024ljf} are not shown here, given that they prove even weaker constraints for heavy mediators and fall outside the parameter space of our interest.

Our results demonstrate that supernova cooling provides the stringent constraints on sub-GeV neutrinophilic dark matter. Compared with indirect detection searches and other cosmological constraints, our bounds on the DM–neutrino reference cross section are more than ten orders of magnitude stronger below a few hundred MeV. The state-of-the-art supernova simulations makes our limits on DM-neutrino interaction reliable and robust. This work highlights the exceptional sensitivity of core-collapse supernovae to feebly interacting dark sectors and strongly motivates future dedicated supernova neutrino observations as a complementary probe of light DM.

\section{SUMMARY AND OUTLOOK}
The extreme temperatures and density environment in core-collapse supernova create a powerful natural laboratory, which is capable of producing light dark-sector particles with masses up to the sub-GeV scale. In particular, the intense burst of neutrinos emitted offers a unique opportunity to  study the interactions between neutrinos and dark matter.

In this work, we investigate neutrino-DM conversion in the supernova using the effective field theory approach via scalar- and vector-type interactions. We set stringent and robust constraints on the
DM-neutrino interaction strength in the sub-GeV DM mass ranges. Importantly, our bounds are much stronger than both current and future indirect detection searches with neutrino detectors, and provide strong complementarity with other cosmological constraints. This work highlights the importance of SN analyses in probing sub-GeV DM scenarios, and it can also be generalized to other types of interactions, or UV models with heavy or light mediators.

\section*{Acknowledgments}
We would like to thank Hans-Thomas Janka for providing the simulation data through the Garching Core Collapse Supernova Archive and many helpful discussions about the simulation data. The authors also acknowledge support from the National Natural Science Foundation of China (12441504).

\begin{appendix}
\section{Simulation Data}
\label{sec:simulationdata}
\begin{figure*}[!htbp]
\centering
\includegraphics[width=\columnwidth]{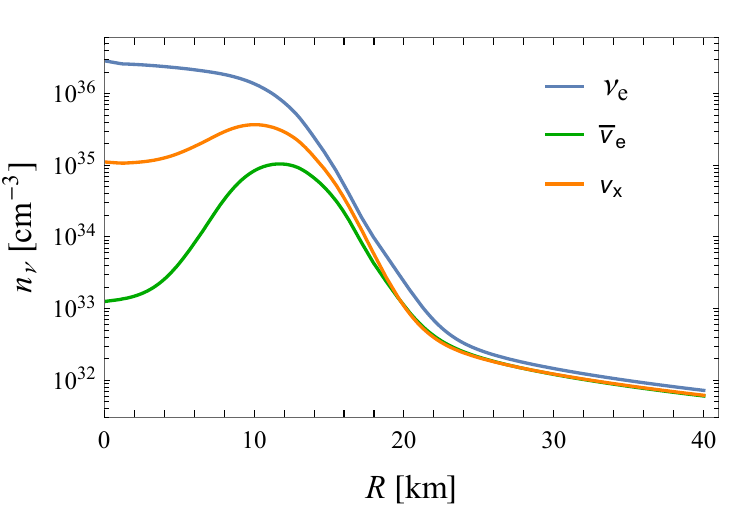}
\includegraphics[width=\columnwidth]{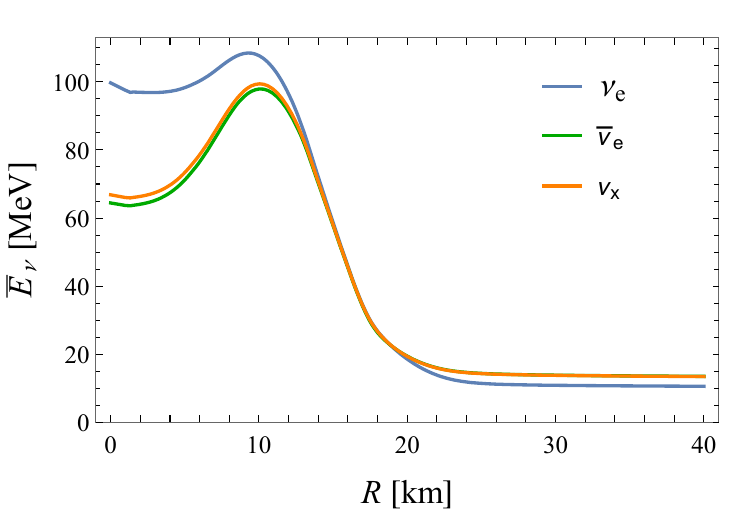}
\includegraphics[width=\columnwidth]{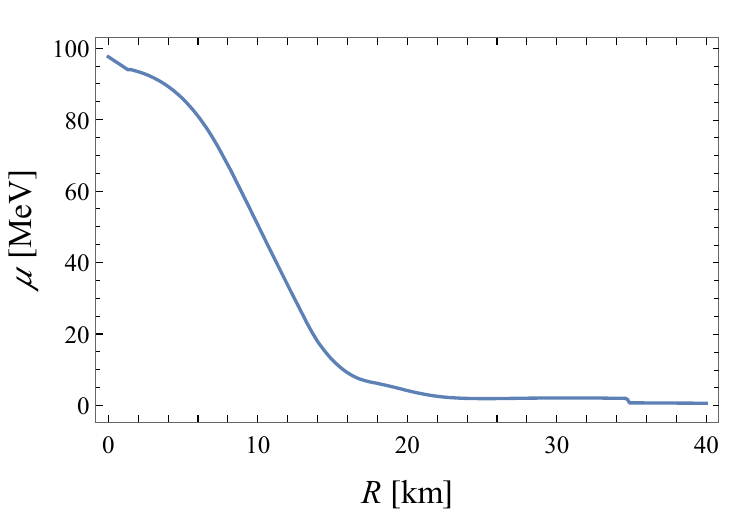}
\includegraphics[width=\columnwidth]{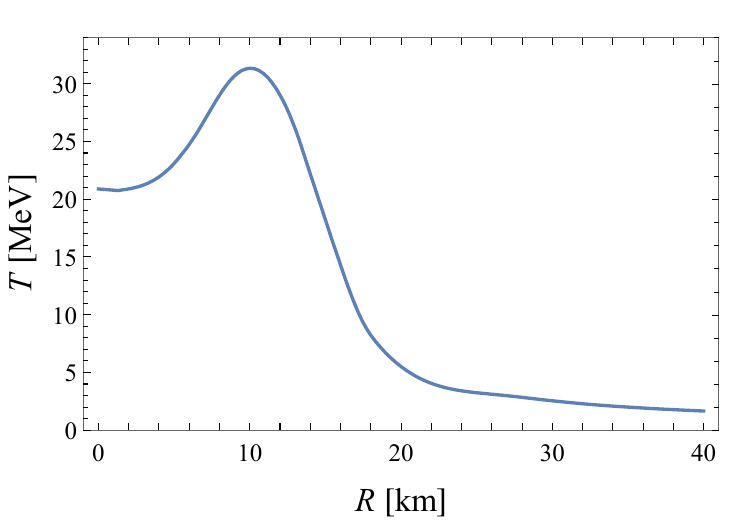}
\caption{The profile of neutrino number density $n_\nu$, mean energy of neutrino $\bar{E}_\nu$, electron neutrino chemical potential $\mu$, and supernova temperature $T$ at one second post core bounce in a CCSN simulation with a 8.8$M_\odot$ progenitor star. The profile of neutrino include $\nu_e$, $\bar{\nu}_e$ and the sum of other neutrino flavors $\nu_x$.}
\label{fig:simulationdata}
\end{figure*}

In this appendix, we show the simulation data of an 8.8$M_\odot$ progenitor star~\cite{Hudepohl:2009tyy} in Fig.~\ref{fig:simulationdata}, including the neutrino number density $n_\nu$, mean energy of neutrino $\bar{E}_\nu$, electron neutrino chemical potential $\mu$, and supernova temperature $T$ at different radii at one second post core bounce, which are as input parameters to calculate the DM energy loss rate presented in this paper.

\end{appendix}

\bibliography{main}

\begin{thebibliography}{50}%
\makeatletter
\providecommand \@ifxundefined [1]{%
 \@ifx{#1\undefined}
}%
\providecommand \@ifnum [1]{%
 \ifnum #1\expandafter \@firstoftwo
 \else \expandafter \@secondoftwo
 \fi
}%
\providecommand \@ifx [1]{%
 \ifx #1\expandafter \@firstoftwo
 \else \expandafter \@secondoftwo
 \fi
}%
\providecommand \natexlab [1]{#1}%
\providecommand \enquote  [1]{``#1''}%
\providecommand \bibnamefont  [1]{#1}%
\providecommand \bibfnamefont [1]{#1}%
\providecommand \citenamefont [1]{#1}%
\providecommand \href@noop [0]{\@secondoftwo}%
\providecommand \href [0]{\begingroup \@sanitize@url \@href}%
\providecommand \@href[1]{\@@startlink{#1}\@@href}%
\providecommand \@@href[1]{\endgroup#1\@@endlink}%
\providecommand \@sanitize@url [0]{\catcode `\\12\catcode `\$12\catcode
  `\&12\catcode `\#12\catcode `\^12\catcode `\_12\catcode `\%12\relax}%
\providecommand \@@startlink[1]{}%
\providecommand \@@endlink[0]{}%
\providecommand \url  [0]{\begingroup\@sanitize@url \@url }%
\providecommand \@url [1]{\endgroup\@href {#1}{\urlprefix }}%
\providecommand \urlprefix  [0]{URL }%
\providecommand \Eprint [0]{\href }%
\providecommand \doibase [0]{https://doi.org/}%
\providecommand \selectlanguage [0]{\@gobble}%
\providecommand \bibinfo  [0]{\@secondoftwo}%
\providecommand \bibfield  [0]{\@secondoftwo}%
\providecommand \translation [1]{[#1]}%
\providecommand \BibitemOpen [0]{}%
\providecommand \bibitemStop [0]{}%
\providecommand \bibitemNoStop [0]{.\EOS\space}%
\providecommand \EOS [0]{\spacefactor3000\relax}%
\providecommand \BibitemShut  [1]{\csname bibitem#1\endcsname}%
\let\auto@bib@innerbib\@empty
\bibitem [{\citenamefont {Mangano}\ \emph {et~al.}(2006)\citenamefont
  {Mangano}, \citenamefont {Melchiorri}, \citenamefont {Serra}, \citenamefont
  {Cooray},\ and\ \citenamefont {Kamionkowski}}]{Mangano:2006mp}%
  \BibitemOpen
  \bibfield  {author} {\bibinfo {author} {\bibfnamefont {G.}~\bibnamefont
  {Mangano}}, \bibinfo {author} {\bibfnamefont {A.}~\bibnamefont {Melchiorri}},
  \bibinfo {author} {\bibfnamefont {P.}~\bibnamefont {Serra}}, \bibinfo
  {author} {\bibfnamefont {A.}~\bibnamefont {Cooray}},\ and\ \bibinfo {author}
  {\bibfnamefont {M.}~\bibnamefont {Kamionkowski}},\ }\bibfield  {title}
  {\bibinfo {title} {{Cosmological bounds on dark matter-neutrino
  interactions}},\ }\href {https://doi.org/10.1103/PhysRevD.74.043517}
  {\bibfield  {journal} {\bibinfo  {journal} {Phys. Rev. D}\ }\textbf {\bibinfo
  {volume} {74}},\ \bibinfo {pages} {043517} (\bibinfo {year} {2006})},\
  \Eprint {https://arxiv.org/abs/astro-ph/0606190} {arXiv:astro-ph/0606190}
  \BibitemShut {NoStop}%
\bibitem [{\citenamefont {Wilkinson}\ \emph {et~al.}(2014)\citenamefont
  {Wilkinson}, \citenamefont {Boehm},\ and\ \citenamefont
  {Lesgourgues}}]{Wilkinson:2014ksa}%
  \BibitemOpen
  \bibfield  {author} {\bibinfo {author} {\bibfnamefont {R.~J.}\ \bibnamefont
  {Wilkinson}}, \bibinfo {author} {\bibfnamefont {C.}~\bibnamefont {Boehm}},\
  and\ \bibinfo {author} {\bibfnamefont {J.}~\bibnamefont {Lesgourgues}},\
  }\bibfield  {title} {\bibinfo {title} {{Constraining Dark Matter-Neutrino
  Interactions using the CMB and Large-Scale Structure}},\ }\href
  {https://doi.org/10.1088/1475-7516/2014/05/011} {\bibfield  {journal}
  {\bibinfo  {journal} {JCAP}\ }\textbf {\bibinfo {volume} {05}},\ \bibinfo
  {pages} {011}},\ \Eprint {https://arxiv.org/abs/1401.7597} {arXiv:1401.7597
  [astro-ph.CO]} \BibitemShut {NoStop}%
\bibitem [{\citenamefont {Brax}\ \emph {et~al.}(2023)\citenamefont {Brax},
  \citenamefont {van~de Bruck}, \citenamefont {Di~Valentino}, \citenamefont
  {Giar{\`e}},\ and\ \citenamefont {Trojanowski}}]{Brax:2023tvn}%
  \BibitemOpen
  \bibfield  {author} {\bibinfo {author} {\bibfnamefont {P.}~\bibnamefont
  {Brax}}, \bibinfo {author} {\bibfnamefont {C.}~\bibnamefont {van~de Bruck}},
  \bibinfo {author} {\bibfnamefont {E.}~\bibnamefont {Di~Valentino}}, \bibinfo
  {author} {\bibfnamefont {W.}~\bibnamefont {Giar{\`e}}},\ and\ \bibinfo
  {author} {\bibfnamefont {S.}~\bibnamefont {Trojanowski}},\ }\bibfield
  {title} {\bibinfo {title} {{Extended analysis of neutrino-dark matter
  interactions with small-scale CMB experiments}},\ }\href
  {https://doi.org/10.1016/j.dark.2023.101321} {\bibfield  {journal} {\bibinfo
  {journal} {Phys. Dark Univ.}\ }\textbf {\bibinfo {volume} {42}},\ \bibinfo
  {pages} {101321} (\bibinfo {year} {2023})},\ \Eprint
  {https://arxiv.org/abs/2305.01383} {arXiv:2305.01383 [astro-ph.CO]}
  \BibitemShut {NoStop}%
\bibitem [{\citenamefont {Boehm}\ \emph {et~al.}(2001)\citenamefont {Boehm},
  \citenamefont {Fayet},\ and\ \citenamefont {Schaeffer}}]{Boehm:2000gq}%
  \BibitemOpen
  \bibfield  {author} {\bibinfo {author} {\bibfnamefont {C.}~\bibnamefont
  {Boehm}}, \bibinfo {author} {\bibfnamefont {P.}~\bibnamefont {Fayet}},\ and\
  \bibinfo {author} {\bibfnamefont {R.}~\bibnamefont {Schaeffer}},\ }\bibfield
  {title} {\bibinfo {title} {{Constraining dark matter candidates from
  structure formation}},\ }\href
  {https://doi.org/10.1016/S0370-2693(01)01060-7} {\bibfield  {journal}
  {\bibinfo  {journal} {Phys. Lett. B}\ }\textbf {\bibinfo {volume} {518}},\
  \bibinfo {pages} {8} (\bibinfo {year} {2001})},\ \Eprint
  {https://arxiv.org/abs/astro-ph/0012504} {arXiv:astro-ph/0012504}
  \BibitemShut {NoStop}%
\bibitem [{\citenamefont {Boehm}\ and\ \citenamefont
  {Schaeffer}(2005)}]{Boehm:2004th}%
  \BibitemOpen
  \bibfield  {author} {\bibinfo {author} {\bibfnamefont {C.}~\bibnamefont
  {Boehm}}\ and\ \bibinfo {author} {\bibfnamefont {R.}~\bibnamefont
  {Schaeffer}},\ }\bibfield  {title} {\bibinfo {title} {{Constraints on dark
  matter interactions from structure formation: Damping lengths}},\ }\href
  {https://doi.org/10.1051/0004-6361:20042238} {\bibfield  {journal} {\bibinfo
  {journal} {Astron. Astrophys.}\ }\textbf {\bibinfo {volume} {438}},\ \bibinfo
  {pages} {419} (\bibinfo {year} {2005})},\ \Eprint
  {https://arxiv.org/abs/astro-ph/0410591} {arXiv:astro-ph/0410591}
  \BibitemShut {NoStop}%
\bibitem [{\citenamefont {Akita}\ and\ \citenamefont
  {Ando}(2023)}]{Akita:2023yga}%
  \BibitemOpen
  \bibfield  {author} {\bibinfo {author} {\bibfnamefont {K.}~\bibnamefont
  {Akita}}\ and\ \bibinfo {author} {\bibfnamefont {S.}~\bibnamefont {Ando}},\
  }\bibfield  {title} {\bibinfo {title} {{Constraints on dark matter-neutrino
  scattering from the Milky-Way satellites and subhalo modeling for dark
  acoustic oscillations}},\ }\href
  {https://doi.org/10.1088/1475-7516/2023/11/037} {\bibfield  {journal}
  {\bibinfo  {journal} {JCAP}\ }\textbf {\bibinfo {volume} {11}},\ \bibinfo
  {pages} {037}},\ \Eprint {https://arxiv.org/abs/2305.01913} {arXiv:2305.01913
  [astro-ph.CO]} \BibitemShut {NoStop}%
\bibitem [{\citenamefont {Heston}\ \emph {et~al.}(2024)\citenamefont {Heston},
  \citenamefont {Horiuchi},\ and\ \citenamefont {Shirai}}]{Heston:2024ljf}%
  \BibitemOpen
  \bibfield  {author} {\bibinfo {author} {\bibfnamefont {S.}~\bibnamefont
  {Heston}}, \bibinfo {author} {\bibfnamefont {S.}~\bibnamefont {Horiuchi}},\
  and\ \bibinfo {author} {\bibfnamefont {S.}~\bibnamefont {Shirai}},\
  }\bibfield  {title} {\bibinfo {title} {{Constraining neutrino-DM interactions
  with Milky~Way dwarf spheroidals and supernova neutrinos}},\ }\href
  {https://doi.org/10.1103/PhysRevD.110.023004} {\bibfield  {journal} {\bibinfo
   {journal} {Phys. Rev. D}\ }\textbf {\bibinfo {volume} {110}},\ \bibinfo
  {pages} {023004} (\bibinfo {year} {2024})},\ \Eprint
  {https://arxiv.org/abs/2402.08718} {arXiv:2402.08718 [hep-ph]} \BibitemShut
  {NoStop}%
\bibitem [{\citenamefont {Dev}\ \emph {et~al.}(2025)\citenamefont {Dev},
  \citenamefont {Kim}, \citenamefont {Sathyan}, \citenamefont {Sinha},\ and\
  \citenamefont {Zhang}}]{Dev:2025tdv}%
  \BibitemOpen
  \bibfield  {author} {\bibinfo {author} {\bibfnamefont {P.~S.~B.}\
  \bibnamefont {Dev}}, \bibinfo {author} {\bibfnamefont {D.}~\bibnamefont
  {Kim}}, \bibinfo {author} {\bibfnamefont {D.}~\bibnamefont {Sathyan}},
  \bibinfo {author} {\bibfnamefont {K.}~\bibnamefont {Sinha}},\ and\ \bibinfo
  {author} {\bibfnamefont {Y.}~\bibnamefont {Zhang}},\ }\bibfield  {title}
  {\bibinfo {title} {{New Constraints on Neutrino-Dark Matter Interactions: A
  Comprehensive Analysis}},\ }\href@noop {} {\  (\bibinfo {year} {2025})},\
  \Eprint {https://arxiv.org/abs/2507.01000} {arXiv:2507.01000 [hep-ph]}
  \BibitemShut {NoStop}%
\bibitem [{\citenamefont {Raffelt}(2008)}]{Raffelt:2006cw}%
  \BibitemOpen
  \bibfield  {author} {\bibinfo {author} {\bibfnamefont {G.~G.}\ \bibnamefont
  {Raffelt}},\ }\bibfield  {title} {\bibinfo {title} {{Astrophysical axion
  bounds}},\ }\href {https://doi.org/10.1007/978-3-540-73518-2_3} {\bibfield
  {journal} {\bibinfo  {journal} {Lect. Notes Phys.}\ }\textbf {\bibinfo
  {volume} {741}},\ \bibinfo {pages} {51} (\bibinfo {year} {2008})},\ \Eprint
  {https://arxiv.org/abs/hep-ph/0611350} {arXiv:hep-ph/0611350} \BibitemShut
  {NoStop}%
\bibitem [{\citenamefont {Hirata}\ \emph {et~al.}(1987)\citenamefont {Hirata}
  \emph {et~al.}}]{Kamiokande-II:1987idp}%
  \BibitemOpen
  \bibfield  {author} {\bibinfo {author} {\bibfnamefont {K.}~\bibnamefont
  {Hirata}} \emph {et~al.} (\bibinfo {collaboration} {Kamiokande-II}),\
  }\bibfield  {title} {\bibinfo {title} {{Observation of a Neutrino Burst from
  the Supernova SN 1987a}},\ }\href
  {https://doi.org/10.1103/PhysRevLett.58.1490} {\bibfield  {journal} {\bibinfo
   {journal} {Phys. Rev. Lett.}\ }\textbf {\bibinfo {volume} {58}},\ \bibinfo
  {pages} {1490} (\bibinfo {year} {1987})}\BibitemShut {NoStop}%
\bibitem [{\citenamefont {Ellis}\ and\ \citenamefont
  {Olive}(1987)}]{Ellis:1987pk}%
  \BibitemOpen
  \bibfield  {author} {\bibinfo {author} {\bibfnamefont {J.~R.}\ \bibnamefont
  {Ellis}}\ and\ \bibinfo {author} {\bibfnamefont {K.~A.}\ \bibnamefont
  {Olive}},\ }\bibfield  {title} {\bibinfo {title} {{Constraints on Light
  Particles From Supernova Sn1987a}},\ }\href
  {https://doi.org/10.1016/0370-2693(87)91710-2} {\bibfield  {journal}
  {\bibinfo  {journal} {Phys. Lett. B}\ }\textbf {\bibinfo {volume} {193}},\
  \bibinfo {pages} {525} (\bibinfo {year} {1987})}\BibitemShut {NoStop}%
\bibitem [{\citenamefont {Raffelt}\ and\ \citenamefont
  {Seckel}(1988)}]{Raffelt:1987yt}%
  \BibitemOpen
  \bibfield  {author} {\bibinfo {author} {\bibfnamefont {G.}~\bibnamefont
  {Raffelt}}\ and\ \bibinfo {author} {\bibfnamefont {D.}~\bibnamefont
  {Seckel}},\ }\bibfield  {title} {\bibinfo {title} {{Bounds on Exotic Particle
  Interactions from SN 1987a}},\ }\href
  {https://doi.org/10.1103/PhysRevLett.60.1793} {\bibfield  {journal} {\bibinfo
   {journal} {Phys. Rev. Lett.}\ }\textbf {\bibinfo {volume} {60}},\ \bibinfo
  {pages} {1793} (\bibinfo {year} {1988})}\BibitemShut {NoStop}%
\bibitem [{\citenamefont {Raffelt}(1996)}]{Raffelt:1996wa}%
  \BibitemOpen
  \bibfield  {author} {\bibinfo {author} {\bibfnamefont {G.~G.}\ \bibnamefont
  {Raffelt}},\ }\href@noop {} {\emph {\bibinfo {title} {{Stars as laboratories
  for fundamental physics}: {The astrophysics of neutrinos, axions, and other
  weakly interacting particles}}}}\ (\bibinfo {year} {1996})\BibitemShut
  {NoStop}%
\bibitem [{\citenamefont {Chang}\ \emph {et~al.}(2018)\citenamefont {Chang},
  \citenamefont {Essig},\ and\ \citenamefont {McDermott}}]{Chang:2018rso}%
  \BibitemOpen
  \bibfield  {author} {\bibinfo {author} {\bibfnamefont {J.~H.}\ \bibnamefont
  {Chang}}, \bibinfo {author} {\bibfnamefont {R.}~\bibnamefont {Essig}},\ and\
  \bibinfo {author} {\bibfnamefont {S.~D.}\ \bibnamefont {McDermott}},\
  }\bibfield  {title} {\bibinfo {title} {{Supernova 1987A Constraints on
  Sub-GeV Dark Sectors, Millicharged Particles, the QCD Axion, and an
  Axion-like Particle}},\ }\href {https://doi.org/10.1007/JHEP09(2018)051}
  {\bibfield  {journal} {\bibinfo  {journal} {JHEP}\ }\textbf {\bibinfo
  {volume} {09}},\ \bibinfo {pages} {051}},\ \Eprint
  {https://arxiv.org/abs/1803.00993} {arXiv:1803.00993 [hep-ph]} \BibitemShut
  {NoStop}%
\bibitem [{\citenamefont {Fiorillo}\ \emph {et~al.}(2023)\citenamefont
  {Fiorillo}, \citenamefont {Raffelt},\ and\ \citenamefont
  {Vitagliano}}]{Fiorillo:2022cdq}%
  \BibitemOpen
  \bibfield  {author} {\bibinfo {author} {\bibfnamefont {D.~F.~G.}\
  \bibnamefont {Fiorillo}}, \bibinfo {author} {\bibfnamefont {G.~G.}\
  \bibnamefont {Raffelt}},\ and\ \bibinfo {author} {\bibfnamefont
  {E.}~\bibnamefont {Vitagliano}},\ }\bibfield  {title} {\bibinfo {title}
  {{Strong Supernova 1987A Constraints on Bosons Decaying to Neutrinos}},\
  }\href {https://doi.org/10.1103/PhysRevLett.131.021001} {\bibfield  {journal}
  {\bibinfo  {journal} {Phys. Rev. Lett.}\ }\textbf {\bibinfo {volume} {131}},\
  \bibinfo {pages} {021001} (\bibinfo {year} {2023})},\ \Eprint
  {https://arxiv.org/abs/2209.11773} {arXiv:2209.11773 [hep-ph]} \BibitemShut
  {NoStop}%
\bibitem [{\citenamefont {Li}\ and\ \citenamefont {Liu}(2025)}]{Li:2025beu}%
  \BibitemOpen
  \bibfield  {author} {\bibinfo {author} {\bibfnamefont {Y.}~\bibnamefont
  {Li}}\ and\ \bibinfo {author} {\bibfnamefont {Z.}~\bibnamefont {Liu}},\
  }\bibfield  {title} {\bibinfo {title} {{Supernova constraints on lepton
  flavor violating ALPs}},\ }\href@noop {} {\  (\bibinfo {year} {2025})},\
  \Eprint {https://arxiv.org/abs/2501.12075} {arXiv:2501.12075 [hep-ph]}
  \BibitemShut {NoStop}%
\bibitem [{\citenamefont {Cappiello}\ \emph {et~al.}(2025)\citenamefont
  {Cappiello}, \citenamefont {Dev},\ and\ \citenamefont
  {Patwardhan}}]{Cappiello:2025tws}%
  \BibitemOpen
  \bibfield  {author} {\bibinfo {author} {\bibfnamefont {C.~V.}\ \bibnamefont
  {Cappiello}}, \bibinfo {author} {\bibfnamefont {P.~S.~B.}\ \bibnamefont
  {Dev}},\ and\ \bibinfo {author} {\bibfnamefont {A.~V.}\ \bibnamefont
  {Patwardhan}},\ }\bibfield  {title} {\bibinfo {title} {{New Supernova
  Constraints on Neutrinophilic Dark Sector}},\ }\href@noop {} {\  (\bibinfo
  {year} {2025})},\ \Eprint {https://arxiv.org/abs/2503.09691}
  {arXiv:2503.09691 [hep-ph]} \BibitemShut {NoStop}%
\bibitem [{\citenamefont {Caputo}\ \emph {et~al.}(2025)\citenamefont {Caputo},
  \citenamefont {Janka}, \citenamefont {Raffelt},\ and\ \citenamefont
  {Yun}}]{Caputo:2025aac}%
  \BibitemOpen
  \bibfield  {author} {\bibinfo {author} {\bibfnamefont {A.}~\bibnamefont
  {Caputo}}, \bibinfo {author} {\bibfnamefont {H.-T.}\ \bibnamefont {Janka}},
  \bibinfo {author} {\bibfnamefont {G.}~\bibnamefont {Raffelt}},\ and\ \bibinfo
  {author} {\bibfnamefont {S.}~\bibnamefont {Yun}},\ }\bibfield  {title}
  {\bibinfo {title} {{Cooling the Shock: New Supernova Constraints on Dark
  Photons}},\ }\href {https://doi.org/10.1103/PhysRevLett.134.151002}
  {\bibfield  {journal} {\bibinfo  {journal} {Phys. Rev. Lett.}\ }\textbf
  {\bibinfo {volume} {134}},\ \bibinfo {pages} {151002} (\bibinfo {year}
  {2025})},\ \Eprint {https://arxiv.org/abs/2502.01731} {arXiv:2502.01731
  [hep-ph]} \BibitemShut {NoStop}%
\bibitem [{\citenamefont {Fiorillo}\ \emph {et~al.}(2024)\citenamefont
  {Fiorillo}, \citenamefont {Raffelt},\ and\ \citenamefont
  {Vitagliano}}]{Fiorillo:2023ytr}%
  \BibitemOpen
  \bibfield  {author} {\bibinfo {author} {\bibfnamefont {D.~F.~G.}\
  \bibnamefont {Fiorillo}}, \bibinfo {author} {\bibfnamefont {G.~G.}\
  \bibnamefont {Raffelt}},\ and\ \bibinfo {author} {\bibfnamefont
  {E.}~\bibnamefont {Vitagliano}},\ }\bibfield  {title} {\bibinfo {title}
  {{Large Neutrino Secret Interactions Have a Small Impact on Supernovae}},\
  }\href {https://doi.org/10.1103/PhysRevLett.132.021002} {\bibfield  {journal}
  {\bibinfo  {journal} {Phys. Rev. Lett.}\ }\textbf {\bibinfo {volume} {132}},\
  \bibinfo {pages} {021002} (\bibinfo {year} {2024})},\ \Eprint
  {https://arxiv.org/abs/2307.15115} {arXiv:2307.15115 [hep-ph]} \BibitemShut
  {NoStop}%
\bibitem [{\citenamefont {Lin}\ \emph {et~al.}(2025)\citenamefont {Lin},
  \citenamefont {Lu},\ and\ \citenamefont {Song}}]{Lin:2025mez}%
  \BibitemOpen
  \bibfield  {author} {\bibinfo {author} {\bibfnamefont {Y.}~\bibnamefont
  {Lin}}, \bibinfo {author} {\bibfnamefont {C.-T.}\ \bibnamefont {Lu}},\ and\
  \bibinfo {author} {\bibfnamefont {N.}~\bibnamefont {Song}},\ }\bibfield
  {title} {\bibinfo {title} {{Supernova cooling from neutrino-devouring dark
  matter}},\ }\href@noop {} {\  (\bibinfo {year} {2025})},\ \Eprint
  {https://arxiv.org/abs/2507.22124} {arXiv:2507.22124 [hep-ph]} \BibitemShut
  {NoStop}%
\bibitem [{\citenamefont {Blennow}\ \emph {et~al.}(2019)\citenamefont
  {Blennow}, \citenamefont {Fernandez-Martinez}, \citenamefont
  {Olivares-Del~Campo}, \citenamefont {Pascoli}, \citenamefont
  {Rosauro-Alcaraz},\ and\ \citenamefont {Titov}}]{Blennow:2019fhy}%
  \BibitemOpen
  \bibfield  {author} {\bibinfo {author} {\bibfnamefont {M.}~\bibnamefont
  {Blennow}}, \bibinfo {author} {\bibfnamefont {E.}~\bibnamefont
  {Fernandez-Martinez}}, \bibinfo {author} {\bibfnamefont {A.}~\bibnamefont
  {Olivares-Del~Campo}}, \bibinfo {author} {\bibfnamefont {S.}~\bibnamefont
  {Pascoli}}, \bibinfo {author} {\bibfnamefont {S.}~\bibnamefont
  {Rosauro-Alcaraz}},\ and\ \bibinfo {author} {\bibfnamefont {A.~V.}\
  \bibnamefont {Titov}},\ }\bibfield  {title} {\bibinfo {title} {{Neutrino
  Portals to Dark Matter}},\ }\href
  {https://doi.org/10.1140/epjc/s10052-019-7060-5} {\bibfield  {journal}
  {\bibinfo  {journal} {Eur. Phys. J. C}\ }\textbf {\bibinfo {volume} {79}},\
  \bibinfo {pages} {555} (\bibinfo {year} {2019})},\ \Eprint
  {https://arxiv.org/abs/1903.00006} {arXiv:1903.00006 [hep-ph]} \BibitemShut
  {NoStop}%
\bibitem [{\citenamefont {Farzan}\ and\ \citenamefont
  {Heeck}(2016)}]{Farzan:2016wym}%
  \BibitemOpen
  \bibfield  {author} {\bibinfo {author} {\bibfnamefont {Y.}~\bibnamefont
  {Farzan}}\ and\ \bibinfo {author} {\bibfnamefont {J.}~\bibnamefont {Heeck}},\
  }\bibfield  {title} {\bibinfo {title} {{Neutrinophilic nonstandard
  interactions}},\ }\href {https://doi.org/10.1103/PhysRevD.94.053010}
  {\bibfield  {journal} {\bibinfo  {journal} {Phys. Rev. D}\ }\textbf {\bibinfo
  {volume} {94}},\ \bibinfo {pages} {053010} (\bibinfo {year} {2016})},\
  \Eprint {https://arxiv.org/abs/1607.07616} {arXiv:1607.07616 [hep-ph]}
  \BibitemShut {NoStop}%
\bibitem [{\citenamefont {Ma}(2006)}]{Ma:2006km}%
  \BibitemOpen
  \bibfield  {author} {\bibinfo {author} {\bibfnamefont {E.}~\bibnamefont
  {Ma}},\ }\bibfield  {title} {\bibinfo {title} {{Verifiable radiative seesaw
  mechanism of neutrino mass and dark matter}},\ }\href
  {https://doi.org/10.1103/PhysRevD.73.077301} {\bibfield  {journal} {\bibinfo
  {journal} {Phys. Rev. D}\ }\textbf {\bibinfo {volume} {73}},\ \bibinfo
  {pages} {077301} (\bibinfo {year} {2006})},\ \Eprint
  {https://arxiv.org/abs/hep-ph/0601225} {arXiv:hep-ph/0601225} \BibitemShut
  {NoStop}%
\bibitem [{\citenamefont {Arhrib}\ \emph {et~al.}(2016)\citenamefont {Arhrib},
  \citenamefont {B{\oe}hm}, \citenamefont {Ma},\ and\ \citenamefont
  {Yuan}}]{Arhrib:2015dez}%
  \BibitemOpen
  \bibfield  {author} {\bibinfo {author} {\bibfnamefont {A.}~\bibnamefont
  {Arhrib}}, \bibinfo {author} {\bibfnamefont {C.}~\bibnamefont {B{\oe}hm}},
  \bibinfo {author} {\bibfnamefont {E.}~\bibnamefont {Ma}},\ and\ \bibinfo
  {author} {\bibfnamefont {T.-C.}\ \bibnamefont {Yuan}},\ }\bibfield  {title}
  {\bibinfo {title} {{Radiative Model of Neutrino Mass with Neutrino
  Interacting MeV Dark Matter}},\ }\href
  {https://doi.org/10.1088/1475-7516/2016/04/049} {\bibfield  {journal}
  {\bibinfo  {journal} {JCAP}\ }\textbf {\bibinfo {volume} {04}},\ \bibinfo
  {pages} {049}},\ \Eprint {https://arxiv.org/abs/1512.08796} {arXiv:1512.08796
  [hep-ph]} \BibitemShut {NoStop}%
\bibitem [{\citenamefont {Babu}\ \emph {et~al.}(2020)\citenamefont {Babu},
  \citenamefont {Dev}, \citenamefont {Jana},\ and\ \citenamefont
  {Thapa}}]{Babu:2019mfe}%
  \BibitemOpen
  \bibfield  {author} {\bibinfo {author} {\bibfnamefont {K.~S.}\ \bibnamefont
  {Babu}}, \bibinfo {author} {\bibfnamefont {P.~S.~B.}\ \bibnamefont {Dev}},
  \bibinfo {author} {\bibfnamefont {S.}~\bibnamefont {Jana}},\ and\ \bibinfo
  {author} {\bibfnamefont {A.}~\bibnamefont {Thapa}},\ }\bibfield  {title}
  {\bibinfo {title} {{Non-Standard Interactions in Radiative Neutrino Mass
  Models}},\ }\href {https://doi.org/10.1007/JHEP03(2020)006} {\bibfield
  {journal} {\bibinfo  {journal} {JHEP}\ }\textbf {\bibinfo {volume} {03}},\
  \bibinfo {pages} {006}},\ \Eprint {https://arxiv.org/abs/1907.09498}
  {arXiv:1907.09498 [hep-ph]} \BibitemShut {NoStop}%
\bibitem [{\citenamefont {Herms}\ \emph {et~al.}(2023)\citenamefont {Herms},
  \citenamefont {Jana}, \citenamefont {K.},\ and\ \citenamefont
  {Saad}}]{Herms:2023cyy}%
  \BibitemOpen
  \bibfield  {author} {\bibinfo {author} {\bibfnamefont {J.}~\bibnamefont
  {Herms}}, \bibinfo {author} {\bibfnamefont {S.}~\bibnamefont {Jana}},
  \bibinfo {author} {\bibfnamefont {V.~P.}\ \bibnamefont {K.}},\ and\ \bibinfo
  {author} {\bibfnamefont {S.}~\bibnamefont {Saad}},\ }\bibfield  {title}
  {\bibinfo {title} {{Light neutrinophilic dark matter from a scotogenic
  model}},\ }\href {https://doi.org/10.1016/j.physletb.2023.138167} {\bibfield
  {journal} {\bibinfo  {journal} {Phys. Lett. B}\ }\textbf {\bibinfo {volume}
  {845}},\ \bibinfo {pages} {138167} (\bibinfo {year} {2023})},\ \Eprint
  {https://arxiv.org/abs/2307.15760} {arXiv:2307.15760 [hep-ph]} \BibitemShut
  {NoStop}%
\bibitem [{\citenamefont {Abdallah}\ \emph {et~al.}(2021)\citenamefont
  {Abdallah}, \citenamefont {Barik}, \citenamefont {Rai},\ and\ \citenamefont
  {Samui}}]{Abdallah:2021npg}%
  \BibitemOpen
  \bibfield  {author} {\bibinfo {author} {\bibfnamefont {W.}~\bibnamefont
  {Abdallah}}, \bibinfo {author} {\bibfnamefont {A.~K.}\ \bibnamefont {Barik}},
  \bibinfo {author} {\bibfnamefont {S.~K.}\ \bibnamefont {Rai}},\ and\ \bibinfo
  {author} {\bibfnamefont {T.}~\bibnamefont {Samui}},\ }\bibfield  {title}
  {\bibinfo {title} {{Search for a light Z' at LHC in a neutrinophilic U(1)
  model}},\ }\href {https://doi.org/10.1103/PhysRevD.104.095031} {\bibfield
  {journal} {\bibinfo  {journal} {Phys. Rev. D}\ }\textbf {\bibinfo {volume}
  {104}},\ \bibinfo {pages} {095031} (\bibinfo {year} {2021})},\ \Eprint
  {https://arxiv.org/abs/2106.01362} {arXiv:2106.01362 [hep-ph]} \BibitemShut
  {NoStop}%
\bibitem [{\citenamefont {Nomura}\ and\ \citenamefont
  {Okada}(2018)}]{Nomura:2017wxf}%
  \BibitemOpen
  \bibfield  {author} {\bibinfo {author} {\bibfnamefont {T.}~\bibnamefont
  {Nomura}}\ and\ \bibinfo {author} {\bibfnamefont {H.}~\bibnamefont {Okada}},\
  }\bibfield  {title} {\bibinfo {title} {{Hidden $U(1)$ gauge symmetry
  realizing a neutrinophilic two-Higgs-doublet model with dark matter}},\
  }\href {https://doi.org/10.1103/PhysRevD.97.075038} {\bibfield  {journal}
  {\bibinfo  {journal} {Phys. Rev. D}\ }\textbf {\bibinfo {volume} {97}},\
  \bibinfo {pages} {075038} (\bibinfo {year} {2018})},\ \Eprint
  {https://arxiv.org/abs/1709.06406} {arXiv:1709.06406 [hep-ph]} \BibitemShut
  {NoStop}%
\bibitem [{\citenamefont {Keil}\ \emph {et~al.}(2003)\citenamefont {Keil},
  \citenamefont {Raffelt},\ and\ \citenamefont {Janka}}]{Keil:2002in}%
  \BibitemOpen
  \bibfield  {author} {\bibinfo {author} {\bibfnamefont {M.~T.}\ \bibnamefont
  {Keil}}, \bibinfo {author} {\bibfnamefont {G.~G.}\ \bibnamefont {Raffelt}},\
  and\ \bibinfo {author} {\bibfnamefont {H.-T.}\ \bibnamefont {Janka}},\
  }\bibfield  {title} {\bibinfo {title} {{Monte Carlo study of supernova
  neutrino spectra formation}},\ }\href {https://doi.org/10.1086/375130}
  {\bibfield  {journal} {\bibinfo  {journal} {Astrophys. J.}\ }\textbf
  {\bibinfo {volume} {590}},\ \bibinfo {pages} {971} (\bibinfo {year}
  {2003})},\ \Eprint {https://arxiv.org/abs/astro-ph/0208035}
  {arXiv:astro-ph/0208035} \BibitemShut {NoStop}%
\bibitem [{\citenamefont {Hudepohl}\ \emph {et~al.}(2010)\citenamefont
  {Hudepohl}, \citenamefont {Muller}, \citenamefont {Janka}, \citenamefont
  {Marek},\ and\ \citenamefont {Raffelt}}]{Hudepohl:2009tyy}%
  \BibitemOpen
  \bibfield  {author} {\bibinfo {author} {\bibfnamefont {L.}~\bibnamefont
  {Hudepohl}}, \bibinfo {author} {\bibfnamefont {B.}~\bibnamefont {Muller}},
  \bibinfo {author} {\bibfnamefont {H.~T.}\ \bibnamefont {Janka}}, \bibinfo
  {author} {\bibfnamefont {A.}~\bibnamefont {Marek}},\ and\ \bibinfo {author}
  {\bibfnamefont {G.~G.}\ \bibnamefont {Raffelt}},\ }\bibfield  {title}
  {\bibinfo {title} {{Neutrino Signal of Electron-Capture Supernovae from Core
  Collapse to Cooling}},\ }\href
  {https://doi.org/10.1103/PhysRevLett.104.251101} {\bibfield  {journal}
  {\bibinfo  {journal} {Phys. Rev. Lett.}\ }\textbf {\bibinfo {volume} {104}},\
  \bibinfo {pages} {251101} (\bibinfo {year} {2010})},\ \bibinfo {note}
  {[Erratum: Phys.Rev.Lett. 105, 249901 (2010)]},\ \Eprint
  {https://arxiv.org/abs/0912.0260} {arXiv:0912.0260 [astro-ph.SR]}
  \BibitemShut {NoStop}%
\bibitem [{\citenamefont {Raffelt}(1990)}]{Raffelt:1990yz}%
  \BibitemOpen
  \bibfield  {author} {\bibinfo {author} {\bibfnamefont {G.~G.}\ \bibnamefont
  {Raffelt}},\ }\bibfield  {title} {\bibinfo {title} {{Astrophysical methods to
  constrain axions and other novel particle phenomena}},\ }\href
  {https://doi.org/10.1016/0370-1573(90)90054-6} {\bibfield  {journal}
  {\bibinfo  {journal} {Phys. Rept.}\ }\textbf {\bibinfo {volume} {198}},\
  \bibinfo {pages} {1} (\bibinfo {year} {1990})}\BibitemShut {NoStop}%
\bibitem [{\citenamefont {Hall}\ \emph {et~al.}(2010)\citenamefont {Hall},
  \citenamefont {Jedamzik}, \citenamefont {March-Russell},\ and\ \citenamefont
  {West}}]{Hall:2009bx}%
  \BibitemOpen
  \bibfield  {author} {\bibinfo {author} {\bibfnamefont {L.~J.}\ \bibnamefont
  {Hall}}, \bibinfo {author} {\bibfnamefont {K.}~\bibnamefont {Jedamzik}},
  \bibinfo {author} {\bibfnamefont {J.}~\bibnamefont {March-Russell}},\ and\
  \bibinfo {author} {\bibfnamefont {S.~M.}\ \bibnamefont {West}},\ }\bibfield
  {title} {\bibinfo {title} {{Freeze-In Production of FIMP Dark Matter}},\
  }\href {https://doi.org/10.1007/JHEP03(2010)080} {\bibfield  {journal}
  {\bibinfo  {journal} {JHEP}\ }\textbf {\bibinfo {volume} {03}},\ \bibinfo
  {pages} {080}},\ \Eprint {https://arxiv.org/abs/0911.1120} {arXiv:0911.1120
  [hep-ph]} \BibitemShut {NoStop}%
\bibitem [{\citenamefont {D'Eramo}\ \emph {et~al.}(2018)\citenamefont
  {D'Eramo}, \citenamefont {Fernandez},\ and\ \citenamefont
  {Profumo}}]{DEramo:2017ecx}%
  \BibitemOpen
  \bibfield  {author} {\bibinfo {author} {\bibfnamefont {F.}~\bibnamefont
  {D'Eramo}}, \bibinfo {author} {\bibfnamefont {N.}~\bibnamefont {Fernandez}},\
  and\ \bibinfo {author} {\bibfnamefont {S.}~\bibnamefont {Profumo}},\
  }\bibfield  {title} {\bibinfo {title} {{Dark Matter Freeze-in Production in
  Fast-Expanding Universes}},\ }\href
  {https://doi.org/10.1088/1475-7516/2018/02/046} {\bibfield  {journal}
  {\bibinfo  {journal} {JCAP}\ }\textbf {\bibinfo {volume} {02}},\ \bibinfo
  {pages} {046}},\ \Eprint {https://arxiv.org/abs/1712.07453} {arXiv:1712.07453
  [hep-ph]} \BibitemShut {NoStop}%
\bibitem [{\citenamefont {Gopalakrishna}\ \emph {et~al.}(2006)\citenamefont
  {Gopalakrishna}, \citenamefont {de~Gouvea},\ and\ \citenamefont
  {Porod}}]{Gopalakrishna:2006kr}%
  \BibitemOpen
  \bibfield  {author} {\bibinfo {author} {\bibfnamefont {S.}~\bibnamefont
  {Gopalakrishna}}, \bibinfo {author} {\bibfnamefont {A.}~\bibnamefont
  {de~Gouvea}},\ and\ \bibinfo {author} {\bibfnamefont {W.}~\bibnamefont
  {Porod}},\ }\bibfield  {title} {\bibinfo {title} {{Right-handed sneutrinos as
  nonthermal dark matter}},\ }\href
  {https://doi.org/10.1088/1475-7516/2006/05/005} {\bibfield  {journal}
  {\bibinfo  {journal} {JCAP}\ }\textbf {\bibinfo {volume} {05}},\ \bibinfo
  {pages} {005}},\ \Eprint {https://arxiv.org/abs/hep-ph/0602027}
  {arXiv:hep-ph/0602027} \BibitemShut {NoStop}%
\bibitem [{\citenamefont {Page}(2007)}]{Page:2007sh}%
  \BibitemOpen
  \bibfield  {author} {\bibinfo {author} {\bibfnamefont {V.}~\bibnamefont
  {Page}},\ }\bibfield  {title} {\bibinfo {title} {{Non-thermal right-handed
  sneutrino dark matter and the Omega(DM)/Omega(b) problem}},\ }\href
  {https://doi.org/10.1088/1126-6708/2007/04/021} {\bibfield  {journal}
  {\bibinfo  {journal} {JHEP}\ }\textbf {\bibinfo {volume} {04}},\ \bibinfo
  {pages} {021}},\ \Eprint {https://arxiv.org/abs/hep-ph/0701266}
  {arXiv:hep-ph/0701266} \BibitemShut {NoStop}%
\bibitem [{\citenamefont {Manzari}\ \emph {et~al.}(2025)\citenamefont
  {Manzari}, \citenamefont {Martin~Camalich}, \citenamefont {Spinner},\ and\
  \citenamefont {Ziegler}}]{Manzari:2025jbc}%
  \BibitemOpen
  \bibfield  {author} {\bibinfo {author} {\bibfnamefont {C.~A.}\ \bibnamefont
  {Manzari}}, \bibinfo {author} {\bibfnamefont {J.}~\bibnamefont
  {Martin~Camalich}}, \bibinfo {author} {\bibfnamefont {J.}~\bibnamefont
  {Spinner}},\ and\ \bibinfo {author} {\bibfnamefont {R.}~\bibnamefont
  {Ziegler}},\ }\bibfield  {title} {\bibinfo {title} {{The SN 1987A Cooling
  Bound on Dark Matter Absorption in Electron Targets}},\ }\href@noop {} {\
  (\bibinfo {year} {2025})},\ \Eprint {https://arxiv.org/abs/2508.00725}
  {arXiv:2508.00725 [hep-ph]} \BibitemShut {NoStop}%
\bibitem [{\citenamefont {Lehmann}\ and\ \citenamefont
  {Profumo}(2020)}]{Lehmann:2020lcv}%
  \BibitemOpen
  \bibfield  {author} {\bibinfo {author} {\bibfnamefont {B.~V.}\ \bibnamefont
  {Lehmann}}\ and\ \bibinfo {author} {\bibfnamefont {S.}~\bibnamefont
  {Profumo}},\ }\bibfield  {title} {\bibinfo {title} {{Cosmology and prospects
  for sub-MeV dark matter in electron recoil experiments}},\ }\href
  {https://doi.org/10.1103/PhysRevD.102.023038} {\bibfield  {journal} {\bibinfo
   {journal} {Phys. Rev. D}\ }\textbf {\bibinfo {volume} {102}},\ \bibinfo
  {pages} {023038} (\bibinfo {year} {2020})},\ \Eprint
  {https://arxiv.org/abs/2002.07809} {arXiv:2002.07809 [hep-ph]} \BibitemShut
  {NoStop}%
\bibitem [{\citenamefont {Kawasaki}\ \emph {et~al.}(2000)\citenamefont
  {Kawasaki}, \citenamefont {Kohri},\ and\ \citenamefont
  {Sugiyama}}]{Kawasaki:2000en}%
  \BibitemOpen
  \bibfield  {author} {\bibinfo {author} {\bibfnamefont {M.}~\bibnamefont
  {Kawasaki}}, \bibinfo {author} {\bibfnamefont {K.}~\bibnamefont {Kohri}},\
  and\ \bibinfo {author} {\bibfnamefont {N.}~\bibnamefont {Sugiyama}},\
  }\bibfield  {title} {\bibinfo {title} {{MeV scale reheating temperature and
  thermalization of neutrino background}},\ }\href
  {https://doi.org/10.1103/PhysRevD.62.023506} {\bibfield  {journal} {\bibinfo
  {journal} {Phys. Rev. D}\ }\textbf {\bibinfo {volume} {62}},\ \bibinfo
  {pages} {023506} (\bibinfo {year} {2000})},\ \Eprint
  {https://arxiv.org/abs/astro-ph/0002127} {arXiv:astro-ph/0002127}
  \BibitemShut {NoStop}%
\bibitem [{\citenamefont {Hannestad}(2004)}]{Hannestad:2004px}%
  \BibitemOpen
  \bibfield  {author} {\bibinfo {author} {\bibfnamefont {S.}~\bibnamefont
  {Hannestad}},\ }\bibfield  {title} {\bibinfo {title} {{What is the lowest
  possible reheating temperature?}},\ }\href
  {https://doi.org/10.1103/PhysRevD.70.043506} {\bibfield  {journal} {\bibinfo
  {journal} {Phys. Rev. D}\ }\textbf {\bibinfo {volume} {70}},\ \bibinfo
  {pages} {043506} (\bibinfo {year} {2004})},\ \Eprint
  {https://arxiv.org/abs/astro-ph/0403291} {arXiv:astro-ph/0403291}
  \BibitemShut {NoStop}%
\bibitem [{\citenamefont {Gondolo}\ and\ \citenamefont
  {Gelmini}(1991)}]{Gondolo:1990dk}%
  \BibitemOpen
  \bibfield  {author} {\bibinfo {author} {\bibfnamefont {P.}~\bibnamefont
  {Gondolo}}\ and\ \bibinfo {author} {\bibfnamefont {G.}~\bibnamefont
  {Gelmini}},\ }\bibfield  {title} {\bibinfo {title} {{Cosmic abundances of
  stable particles: Improved analysis}},\ }\href
  {https://doi.org/10.1016/0550-3213(91)90438-4} {\bibfield  {journal}
  {\bibinfo  {journal} {Nucl. Phys. B}\ }\textbf {\bibinfo {volume} {360}},\
  \bibinfo {pages} {145} (\bibinfo {year} {1991})}\BibitemShut {NoStop}%
\bibitem [{\citenamefont {Cline}\ and\ \citenamefont
  {Puel}(2023)}]{Cline:2023tkp}%
  \BibitemOpen
  \bibfield  {author} {\bibinfo {author} {\bibfnamefont {J.~M.}\ \bibnamefont
  {Cline}}\ and\ \bibinfo {author} {\bibfnamefont {M.}~\bibnamefont {Puel}},\
  }\bibfield  {title} {\bibinfo {title} {{NGC 1068 constraints on neutrino-dark
  matter scattering}},\ }\href {https://doi.org/10.1088/1475-7516/2023/06/004}
  {\bibfield  {journal} {\bibinfo  {journal} {JCAP}\ }\textbf {\bibinfo
  {volume} {06}},\ \bibinfo {pages} {004}},\ \Eprint
  {https://arxiv.org/abs/2301.08756} {arXiv:2301.08756 [hep-ph]} \BibitemShut
  {NoStop}%
\bibitem [{\citenamefont {Bellini}\ \emph {et~al.}(2011)\citenamefont {Bellini}
  \emph {et~al.}}]{Borexino:2010zht}%
  \BibitemOpen
  \bibfield  {author} {\bibinfo {author} {\bibfnamefont {G.}~\bibnamefont
  {Bellini}} \emph {et~al.} (\bibinfo {collaboration} {Borexino}),\ }\bibfield
  {title} {\bibinfo {title} {{Study of solar and other unknown anti-neutrino
  fluxes with Borexino at LNGS}},\ }\href
  {https://doi.org/10.1016/j.physletb.2010.12.030} {\bibfield  {journal}
  {\bibinfo  {journal} {Phys. Lett. B}\ }\textbf {\bibinfo {volume} {696}},\
  \bibinfo {pages} {191} (\bibinfo {year} {2011})},\ \Eprint
  {https://arxiv.org/abs/1010.0029} {arXiv:1010.0029 [hep-ex]} \BibitemShut
  {NoStop}%
\bibitem [{\citenamefont {Bellini}\ \emph {et~al.}(2010)\citenamefont {Bellini}
  \emph {et~al.}}]{Borexino:2010dli}%
  \BibitemOpen
  \bibfield  {author} {\bibinfo {author} {\bibfnamefont {G.}~\bibnamefont
  {Bellini}} \emph {et~al.} (\bibinfo {collaboration} {Borexino}),\ }\bibfield
  {title} {\bibinfo {title} {{Observation of Geo-Neutrinos}},\ }\href
  {https://doi.org/10.1016/j.physletb.2010.03.051} {\bibfield  {journal}
  {\bibinfo  {journal} {Phys. Lett. B}\ }\textbf {\bibinfo {volume} {687}},\
  \bibinfo {pages} {299} (\bibinfo {year} {2010})},\ \Eprint
  {https://arxiv.org/abs/1003.0284} {arXiv:1003.0284 [hep-ex]} \BibitemShut
  {NoStop}%
\bibitem [{\citenamefont {Gando}\ \emph {et~al.}(2012)\citenamefont {Gando}
  \emph {et~al.}}]{KamLAND:2011bnd}%
  \BibitemOpen
  \bibfield  {author} {\bibinfo {author} {\bibfnamefont {A.}~\bibnamefont
  {Gando}} \emph {et~al.} (\bibinfo {collaboration} {KamLAND}),\ }\bibfield
  {title} {\bibinfo {title} {{A study of extraterrestrial antineutrino sources
  with the KamLAND detector}},\ }\href
  {https://doi.org/10.1088/0004-637X/745/2/193} {\bibfield  {journal} {\bibinfo
   {journal} {Astrophys. J.}\ }\textbf {\bibinfo {volume} {745}},\ \bibinfo
  {pages} {193} (\bibinfo {year} {2012})},\ \Eprint
  {https://arxiv.org/abs/1105.3516} {arXiv:1105.3516 [astro-ph.HE]}
  \BibitemShut {NoStop}%
\bibitem [{\citenamefont {Gando}\ \emph {et~al.}(2013)\citenamefont {Gando}
  \emph {et~al.}}]{KamLAND:2013rgu}%
  \BibitemOpen
  \bibfield  {author} {\bibinfo {author} {\bibfnamefont {A.}~\bibnamefont
  {Gando}} \emph {et~al.} (\bibinfo {collaboration} {KamLAND}),\ }\bibfield
  {title} {\bibinfo {title} {{Reactor On-Off Antineutrino Measurement with
  KamLAND}},\ }\href {https://doi.org/10.1103/PhysRevD.88.033001} {\bibfield
  {journal} {\bibinfo  {journal} {Phys. Rev. D}\ }\textbf {\bibinfo {volume}
  {88}},\ \bibinfo {pages} {033001} (\bibinfo {year} {2013})},\ \Eprint
  {https://arxiv.org/abs/1303.4667} {arXiv:1303.4667 [hep-ex]} \BibitemShut
  {NoStop}%
\bibitem [{\citenamefont {Cravens}\ \emph {et~al.}(2008)\citenamefont {Cravens}
  \emph {et~al.}}]{Super-Kamiokande:2008ecj}%
  \BibitemOpen
  \bibfield  {author} {\bibinfo {author} {\bibfnamefont {J.~P.}\ \bibnamefont
  {Cravens}} \emph {et~al.} (\bibinfo {collaboration} {Super-Kamiokande}),\
  }\bibfield  {title} {\bibinfo {title} {{Solar neutrino measurements in
  Super-Kamiokande-II}},\ }\href {https://doi.org/10.1103/PhysRevD.78.032002}
  {\bibfield  {journal} {\bibinfo  {journal} {Phys. Rev. D}\ }\textbf {\bibinfo
  {volume} {78}},\ \bibinfo {pages} {032002} (\bibinfo {year} {2008})},\
  \Eprint {https://arxiv.org/abs/0803.4312} {arXiv:0803.4312 [hep-ex]}
  \BibitemShut {NoStop}%
\bibitem [{\citenamefont {Abe}\ \emph {et~al.}(2011)\citenamefont {Abe} \emph
  {et~al.}}]{Super-Kamiokande:2010tar}%
  \BibitemOpen
  \bibfield  {author} {\bibinfo {author} {\bibfnamefont {K.}~\bibnamefont
  {Abe}} \emph {et~al.} (\bibinfo {collaboration} {Super-Kamiokande}),\
  }\bibfield  {title} {\bibinfo {title} {{Solar neutrino results in
  Super-Kamiokande-III}},\ }\href {https://doi.org/10.1103/PhysRevD.83.052010}
  {\bibfield  {journal} {\bibinfo  {journal} {Phys. Rev. D}\ }\textbf {\bibinfo
  {volume} {83}},\ \bibinfo {pages} {052010} (\bibinfo {year} {2011})},\
  \Eprint {https://arxiv.org/abs/1010.0118} {arXiv:1010.0118 [hep-ex]}
  \BibitemShut {NoStop}%
\bibitem [{\citenamefont {Bell}\ \emph {et~al.}(2020)\citenamefont {Bell},
  \citenamefont {Dolan},\ and\ \citenamefont {Robles}}]{Bell:2020rkw}%
  \BibitemOpen
  \bibfield  {author} {\bibinfo {author} {\bibfnamefont {N.~F.}\ \bibnamefont
  {Bell}}, \bibinfo {author} {\bibfnamefont {M.~J.}\ \bibnamefont {Dolan}},\
  and\ \bibinfo {author} {\bibfnamefont {S.}~\bibnamefont {Robles}},\
  }\bibfield  {title} {\bibinfo {title} {{Searching for Sub-GeV Dark Matter in
  the Galactic Centre using Hyper-Kamiokande}},\ }\href
  {https://doi.org/10.1088/1475-7516/2020/09/019} {\bibfield  {journal}
  {\bibinfo  {journal} {JCAP}\ }\textbf {\bibinfo {volume} {09}},\ \bibinfo
  {pages} {019}},\ \Eprint {https://arxiv.org/abs/2005.01950} {arXiv:2005.01950
  [hep-ph]} \BibitemShut {NoStop}%
\bibitem [{\citenamefont {An}\ \emph {et~al.}(2016)\citenamefont {An} \emph
  {et~al.}}]{JUNO:2015zny}%
  \BibitemOpen
  \bibfield  {author} {\bibinfo {author} {\bibfnamefont {F.}~\bibnamefont {An}}
  \emph {et~al.} (\bibinfo {collaboration} {JUNO}),\ }\bibfield  {title}
  {\bibinfo {title} {{Neutrino Physics with JUNO}},\ }\href
  {https://doi.org/10.1088/0954-3899/43/3/030401} {\bibfield  {journal}
  {\bibinfo  {journal} {J. Phys. G}\ }\textbf {\bibinfo {volume} {43}},\
  \bibinfo {pages} {030401} (\bibinfo {year} {2016})},\ \Eprint
  {https://arxiv.org/abs/1507.05613} {arXiv:1507.05613 [physics.ins-det]}
  \BibitemShut {NoStop}%
\bibitem [{\citenamefont {Arg{\"u}elles}\ \emph {et~al.}(2021)\citenamefont
  {Arg{\"u}elles}, \citenamefont {Diaz}, \citenamefont {Kheirandish},
  \citenamefont {Olivares-Del-Campo}, \citenamefont {Safa},\ and\ \citenamefont
  {Vincent}}]{Arguelles:2019ouk}%
  \BibitemOpen
  \bibfield  {author} {\bibinfo {author} {\bibfnamefont {C.~A.}\ \bibnamefont
  {Arg{\"u}elles}}, \bibinfo {author} {\bibfnamefont {A.}~\bibnamefont {Diaz}},
  \bibinfo {author} {\bibfnamefont {A.}~\bibnamefont {Kheirandish}}, \bibinfo
  {author} {\bibfnamefont {A.}~\bibnamefont {Olivares-Del-Campo}}, \bibinfo
  {author} {\bibfnamefont {I.}~\bibnamefont {Safa}},\ and\ \bibinfo {author}
  {\bibfnamefont {A.~C.}\ \bibnamefont {Vincent}},\ }\bibfield  {title}
  {\bibinfo {title} {{Dark matter annihilation to neutrinos}},\ }\href
  {https://doi.org/10.1103/RevModPhys.93.035007} {\bibfield  {journal}
  {\bibinfo  {journal} {Rev. Mod. Phys.}\ }\textbf {\bibinfo {volume} {93}},\
  \bibinfo {pages} {035007} (\bibinfo {year} {2021})},\ \Eprint
  {https://arxiv.org/abs/1912.09486} {arXiv:1912.09486 [hep-ph]} \BibitemShut
  {NoStop}%
\end{thebibliography}%

\end{document}